\documentclass[10pt,twocolumn,english]{revtex4}
\usepackage[T1]{fontenc}
\usepackage[latin9]{inputenc}
\usepackage[a4paper]{geometry}
\usepackage{amsmath}
\usepackage{amssymb}
\usepackage{graphicx}

\makeatletter

\providecommand{\tabularnewline}{\\}

\@ifundefined{textcolor}{}
{%
 \definecolor{BLACK}{gray}{0}
 \definecolor{WHITE}{gray}{1}
 \definecolor{RED}{rgb}{1,0,0}
 \definecolor{GREEN}{rgb}{0,1,0}
 \definecolor{BLUE}{rgb}{0,0,1}
 \definecolor{CYAN}{cmyk}{1,0,0,0}
 \definecolor{MAGENTA}{cmyk}{0,1,0,0}
 \definecolor{YELLOW}{cmyk}{0,0,1,0}
}

\makeatother

\usepackage{babel}
\begin{document}

\title{Consistent closure of renormalization group flow equations in quantum gravity}

\author{Alessandro Codello$^a$, Giulio D'Odorico$^b$ and Carlo Pagani$^{a,c}$\\
$^a$\emph{SISSA},
\emph{via Bonomea 265,} \emph{I-34136} \emph{Trieste, Italy}\\
$^b$\emph{IMAPP}, \emph{Radboud University},
\emph{Heyendaalseweg 135, 6525 AJ, Nijmegen, The Netherlands}\\
$^c$\emph{INFN, Sezione di Trieste, Italy}
}
\begin{abstract}
We construct a consistent closure for the beta functions of the cosmological and Newton's constants by evaluating the influence that the anomalous dimensions of the fluctuating metric and ghost fields have on their renormalization group flow.
In this generalized framework we confirm the presence of an UV attractive non--Gaussian fixed--point, which we find characterized by real critical exponents.
%
Our closure method is general and can be applied systematically to more general truncations of the gravitational effective average action.
\end{abstract}
\maketitle

\section{Introduction}

A promising approach to quantum gravity is the Asymptotic Safety scenario, first proposed by Weinberg \cite{Weinberg},
which aims to describe quantum gravity within the framework of quantum field theory.
As is well known, the quantum field theory of gravity based on the Einstein--Hilbert action is perturbatively non--renormalizable \cite{Hooft};
the Asymptotic Safety scenario suggests instead that the theory is non--perturbatively renormalizable at
a non--Gaussian ultraviolet (UV) fixed--point of the renormalization group (RG) flow.
To inquire if a theory is renormalizable in a non--perturbative way one needs non-perturbative tools.
One of these tools is the exact functional RG equation satisfied by the (background) effective average action (EAA),
first derived in the context of quantum gravity by Reuter \cite{Reuter:1996cp}.
%
So far, various applications of the EAA formalism to the problem of quantum gravity \cite{Reuter:1996cp,Codello:2008vh}
have supported the Asymptotic Safety scenario. Evidence has been found for the existence of a non--Gaussian fixed--point with a finite dimensional UV--critical surface \cite{Codello:2007bd}.
%
%
All previous applications of the EAA formalism to quantum gravity were based on the specific closure of the beta functions first proposed in \cite{Reuter:1996cp}; in this letter we will propose a more consistent approach that accounts for the non--trivial influence that the anomalous dimensions of the fluctuating metric and ghost fields have on the RG flow of the cosmological and Newton's constants.
Our results show that even in this enlarged framework, these two couplings are characterized by a non--Gaussian fixed point, thus attesting the robustness of Asymptotic Safety to this generalization.

The application of functional renormalization group techniques to theories characterized by local
symmetries requires overcoming the problem of performing the coarse--graining procedure in a covariant way.
A solution to this problem comes from the combination of the EAA and the background field formalisms \cite{Reuter:1993kw}.
The preservation of gauge invariance along the flow comes at the cost of enlarging
theory space to include invariants constructed with both background and fluctuating fields.
This defines the background EAA, which becomes a functional of these two fields invariant under physical and background gauge transformations.
%
In the case of gravity, the metric $g_{\mu \nu}$ is split into a background $\bar{g}_{\mu \nu}$ and a fluctuation
$h_{\mu \nu}$ in the following way:
\begin{equation}
g_{\mu\nu} = \bar{g}_{\mu\nu} + \sqrt{32\pi G_{k}} \, h_{\mu\nu}\,.   \label{gravity_Q_3}
\end{equation}
We have re--scaled the fluctuating metric so that the combination $ \sqrt{32\pi G_{k}}$ acts as the gravitational coupling, $G_k$ being the scale dependent Newton's constant (see eq. (\ref{gravity_Q_2}) in the next section).
In this way a gravitational vertex with $n$--legs is accompanied by a factor $(\sqrt{32\pi G_{k}})^{n-2}$.
In the construction of the background EAA one introduces, in the path integral, source, gauge--fixing and cutoff terms;
in these the background and fluctuating metric do not appear via their sum $g_{\mu \nu}$.
As a consequence, the RG flow generates invariants which depend on $\bar{g}_{\mu \nu}$ and $h_{\mu \nu}$ separately.
In this formalism, only couplings of invariants which are functionals of $g_{\mu \nu}$ alone are interpreted as physical,
while the couplings of the other invariants are seen as unphysical, even if their influence on the flow of physical couplings is non--trivial.
The study of this influence is the main goal of this letter.
%

The background EAA for gravity splits as follows: 
\begin{eqnarray}
\Gamma_{k}[\varphi;\bar{g}]=\bar{\Gamma}_{k}[\bar{g}+h]+\hat{\Gamma}_{k}[\varphi;\bar{g}]\,,	\label{EAA}
\end{eqnarray}
where $\varphi=(h_{\mu\nu},\bar{C}_\mu,C^\nu)$ is the fluctuating multiplet comprising the fluctuating metric and the ghost fields.
In (\ref{EAA}) we defined the gauge invariant EAA $\bar{\Gamma}_{k}[\bar{g}]$ and the remainder EAA  $\hat{\Gamma}_{k}[0;\bar{g}]$.
For consistency we must have $\bar{\Gamma}_{k}[\bar{g}]\equiv\Gamma_{k}[0;\bar{g}]$ and $\hat{\Gamma}_{k}[0;\bar{g}]=0$, with the first invariant under physical diffeomorphisms and the second invariant under combined physical and background diffeomorphisms.
%
The main virtue of the background EAA for gravity is that it satisfies an exact equation \cite{Reuter:1996cp}:
\begin{eqnarray}
\partial_{t}\Gamma_{k}[\varphi;\bar{g}]=\frac{1}{2}\mathrm{Tr}\left(\Gamma_{k}^{(2;0)}[\varphi;\bar{g}]+R_{k}[\bar{g}]\right)^{-1}\,\partial_{t}R_{k}[\bar{g}]\,,	\label{ERGE}
\end{eqnarray}
which defines a mathematically consistent RG flow, i.e. UV and IR finite, despite the theory being perturbatively non--renormalizable:
this is the non-perturbative tool that is used to inquire if gravity is asymptotically safe.

It is important to note that the RG flow described by (\ref{ERGE}) is driven by the Hessian of the background EAA taken with respect to the fluctuating multiplet $\varphi$,
thus the RG flow equation for $\bar{\Gamma}_{k}[\bar{g}]$, resulting from setting $\varphi=0$ in (\ref{ERGE}), is \emph{not closed} since its rhs depends also on $\hat{\Gamma}_k[\varphi;\bar{g}]$.
This fact forces us to consider the flow of the full $\Gamma_{k}[\varphi,\bar{g}]$ instead of only that of $\bar{\Gamma}_{k}[\bar{g}]$, which we would like to consider the physically interesting one.
It is thus of fundamental importance to develop a systematic way to treat truncations of the full background EAA depending
on both background and fluctuating fields, in order to consistently close the RG flow of the gauge invariant part of the EAA.
%
In this letter we will apply to the Einstein--Hilbert truncation the technique developed in \cite{Codello2013} for this purpose.
We will show how the RG flow of the cosmological and Newton's constants
can be consistently closed by independently evaluating
the anomalous dimensions of the fluctuating metric and ghost fields.

\section{Einstein--Hilbert truncation}

In actual applications one typically makes an ansatz for the background EAA $\Gamma_k[\varphi , \bar{g}]$;
this means that theory space is truncated to some chosen functional and one hopes that this subspace
is complete enough to describe the flow in an approximate, but yet non-perturbative, way.
%
Our truncation ansatz for the gauge invariant part of the EAA will be the RG improved
version of the Einstein--Hilbert action, where both the
cosmological and Newton's constants become scale dependent quantities:
\begin{equation}
\bar{\Gamma}_{k}[g]=\frac{1}{16\pi G_{k}}\int d^{d}x\sqrt{g}\left(2\Lambda_{k}-R\right)\,. \label{gravity_Q_2}
\end{equation}
Quantum fluctuations are responsible for the anomalous scaling of the fields; this fact is accounted
for by introducing scale dependent wave--function renormalization constants for all the fluctuating
fields present in the theory; in our case we redefine the fluctuating metric and the ghost fields according to:
\begin{equation}
h_{\mu\nu}\rightarrow Z_{h,k}^{1/2}h_{\mu\nu}\quad\bar{C}_{\mu}\rightarrow Z_{C,k}^{1/2}\bar{C}_{\mu}\quad C^{\nu}\rightarrow Z_{C,k}^{1/2}C^{\nu}\,. \label{RF}
\end{equation}
The scale derivative of the logarithm of the wave--function renormalization constants defines the relative anomalous dimensions:
\begin{equation}
\eta_{h,k}=-\partial_{t}\log Z_{h,k}\quad\quad\eta_{C,k}=-\partial_{t}\log Z_{C,k}\,.  \label{AD}
\end{equation} 
%
Next, we need to make an ansatz for the remainder functional $\hat{\Gamma}_k[\varphi,\bar{g}]$.
We will consider the simplest non--trivial case comprised by the classical background gauge--fixing and ghost actions (these can be found in \cite{Codello:2008vh}).
We work with the gauge condition $\alpha=\beta=1$.


There are different possible cutoff choices which can be thought of as the freedom we have in setting
up our coarse--graining procedure. In the nomenclature of \cite{Codello:2008vh},
we will present the results for the type Ia cutoff in order to compare to previous findings \cite{Groh:2010ta}.
The type Ia cutoff is characterized by having as cutoff operators the covariant Laplacians
in both the gravitational and ghost sectors.

\section{Beta functions}

To obtain the beta functions of the physical couplings one computes the Hessian of the background EAA with respect
to the fluctuating fields $\varphi$, inserts it into the rhs of the RG flow equation (\ref{ERGE}) and then sets $\varphi=0$.
The trace on the rhs of (\ref{ERGE}) can then be expanded in terms of invariants of
the background metric using heat kernel techniques in a standard way \cite{Codello:2008vh}.

After introducing dimensionless cosmological and Newton's constants,
$\tilde{\Lambda}_{k}=k^{-2}\Lambda_{k}$ and $\tilde{G}_{k}=k^{d-2}G_{k}$,
one finds the following general form for the beta functions:
\begin{eqnarray}
\partial_{t}\tilde{\Lambda}_{k}=-2\tilde{\Lambda}_{k}\nonumber\qquad\qquad\qquad\qquad\qquad\qquad\qquad\qquad\\
+\left[A_{d}(\tilde{\Lambda}_k)+C_{d}(\tilde{\Lambda}_k)\,\eta_{h,k}+E_{d}(\tilde{\Lambda}_k)\,\eta_{C,k}\right]\tilde{G}_{k}\nonumber\\
\partial_{t}\tilde{G}_{k}=(d-2)\tilde{G}_{k}\nonumber\qquad\qquad\qquad\qquad\qquad\qquad\qquad\,\\
+\left[B_{d}(\tilde{\Lambda}_k)+D_{d}(\tilde{\Lambda}_k)\,\eta_{h,k}+F_{d}(\tilde{\Lambda}_k)\,\eta_{C,k}\right]\tilde{G}_{k}^{2}\,, \label{BF}
\end{eqnarray}
where $A_{d},B_{d},C_{d},D_{d},E_{d},F_{d}$ are functions of the dimensionless
cosmological constant $\tilde{\Lambda}_k$ and their specific form 
depends on both the cutoff type and cutoff shape function (i.e. the smearing function in the cutoff action that dictates how slow modes are suppressed).
The explicit form of the functions $A_{d},B_{d},C_{d},D_{d}$ can be found in \cite{Codello:2008vh},
while the functions $E_{d},F_{d}$ can be extracted from \cite{Groh:2010ta}. 
The beta functions (\ref{BF}) constitute the basis of the EAA approach to quantum gravity and were first derived in \cite{Reuter:1996cp}.
Note that they are valid for $d\ge2$.
%

The beta functions (\ref{BF}) for the physical couplings $\Lambda_{k}$ and $G_{k}$ do not form a closed system of ODEs 
because of the presence, on the rhs, of the anomalous dimensions $\eta_{h,k}$ and $\eta_{C,k}$.
This reflects the fact, noticed previously, 
that the RG flow equation for $\bar{\Gamma}_{k}[\bar{g}]$ is not closed, but depends also on $\hat{\Gamma}_k[\varphi;\bar{g}]$, and
this forces us to consider the flow of the full background EAA to find a consistent closure for these beta functions. 
%
%
Before presenting our solution to this problem, we will review the approaches available in the literature.
\begin{table}
\begin{centering}
\begin{tabular}{|c|c|c|c|c|c|c|}
\hline 
 & {\small $\tilde{\Lambda}_{*}$} & {\small $\tilde{G}_{*}$} & {\small $\theta'\pm i\theta''$} & {\small $\tilde{\Lambda}_{*}\tilde{G}_{*}$} & {\small ${\eta_{h,*}}$} & {\small ${\eta_{C,*}}$}\tabularnewline
\hline 
{\small One--loop} & {\small $0.121$} & {\small $1.172$} & {\small $-1.868\pm1.398i$} & {\small $0.142$} & {\small $0$} & {\small $0$}\tabularnewline
\hline 
{\small \cite{Reuter:1996cp}} & {\small $0.193$} & {\small $0.707$} & {\small $-1.475\pm3.043i$} & {\small $0.137$} & {\small $-2$} & {\small $0$}\tabularnewline
\hline 
{\small \cite{Groh:2010ta}} & {\small $0.135$} & {\small $0.859$} & {\small $-1.774\pm1.935i$} & {\small $0.116$} & {\small $-2$} & {\small $-1.8$}\tabularnewline
\hline 
{\small This work} & {\small $-0.062$} & {\small $1.617$} & {\small $-4.119,-1.338$} & {\small $-0.100$} & {\small $0.686$} & {\small $-1.356$}\tabularnewline
\hline 
\end{tabular}
\par\end{centering}
\caption{Fixed--points and critical exponents for the various closures of the
beta functions of $\Lambda_{k}$ and $G_{k}$, in $d=4$.}
\end{table}

\section{Closing the flow equations}

The first way in which one can close the beta functions (\ref{BF}) 
is the trivial one where one sets $\eta_{h,k}=\eta_{C,k}=0$.
This amounts to a one--loop approximation. Within this approximation
only the first terms inside the parenthesis of (\ref{BF}) are retained \cite{Codello:2008vh}: 
\begin{eqnarray}
\partial_{t}\tilde{\Lambda}_{k}&=&-2\tilde{\Lambda}_{k}+A_{d}(\tilde{\Lambda}_k)\tilde{G}_{k}\nonumber\\
\partial_{t}\tilde{G}_{k}&=&(d-2)\tilde{G}_{k}+B_{d}(\tilde{\Lambda}_k)\tilde{G}_{k}^{2}\,. \label{BF1L}
\end{eqnarray}
%
A similar one--loop flow is generated by matter interactions and becomes physically significant when the number of matter fields is large \cite{Percacci:2005wu}.
%
In two dimensions the beta function for Newton's constant is universal and one indeed finds $B_2(0)=-\frac{38}{3}$ for any cutoff specification \cite{Codello:2008vh}.

The second closure method is the RG improvement
adopted in most previous studies \cite{Reuter:1996cp,Dou:1997fg,Codello:2008vh}.
The system (\ref{BF}) is closed by imposing the following relations:
\begin{equation}
\eta_{h,k}=\frac{\partial_t G_k}{G_k}=2-d+\frac{\partial_t \tilde{G}_k}{\tilde{G}_k}\qquad\quad \eta_{C,k}=0\,.\label{std}
\end{equation}
The identification in (\ref{std}) implies a non--trivial, but difficult to interpret, RG improvement of the beta functions.
We will call this procedure the \emph{standard} RG improvement of the beta functions (\ref{BF}).
Note also that in this way we are imposing $\eta_{h,*}=2-d$ at any non-Gaussian fixed--point \cite{Rosten:2011mf}.
The beta functions obtained in this way are exactly those first obtained in \cite{Reuter:1996cp}.
Inserting (\ref{std}) in (\ref{BF}) and solving for $\partial_t \tilde{G}_k$ gives:
\begin{eqnarray}
\partial_{t}\tilde{\Lambda}_{k}&=&-2\tilde{\Lambda}_{k}+A_{d}(\tilde{\Lambda}_k)\tilde{G}_k
+\frac{B_{d}(\tilde{\Lambda}_k) C_{d}(\tilde{\Lambda}_k)}{1-D_{d}(\tilde{\Lambda}_k)\tilde{G}_{k}}\tilde{G}_{k}^2\nonumber\\
\partial_{t}\tilde{G}_{k}&=&(d-2)\tilde{G}_{k}
+\frac{B_{d}(\tilde{\Lambda}_k)\tilde{G}_{k}^{2}}{1-D_{d}(\tilde{\Lambda}_k)\tilde{G}_{k}} \,. \quad\quad \label{BFstd}
\end{eqnarray}
Note that these are rational functions of both $\tilde{G}_{k}$
and $\tilde{\Lambda}_{k}$; this can be interpreted as a re-summation
of an infinite number of perturbative diagrams implemented by the
RG improvement implied by (\ref{std}).
%

In $d=4$, the beta functions (\ref{BFstd}) have a non--Gaussian fixed--point for the values of $\tilde{\Lambda}_{*}$ and $\tilde{G}_{*}$
reported in Table 1.
%
The non--Gaussian fixed--point is UV attractive in both directions;
thus, within this truncation, quantum gravity is asymptotically safe.
The stability matrix has a pair of complex conjugated critical exponents with negative real part.
These are also reported in Table 1.
Here we follow the convention of \cite{Codello:2008vh}
that a negative value for the critical exponent implies that the relative
eigendirection is UV attractive. For more details on this point see \cite{Codello:2008vh}.
To guarantee predictivity we still need to show that the UV critical surface is finite dimensional;
to do this we need to enlarge our truncation and see if we find operators with repulsive UV directions at the non--Gaussian fixed--point.
Evidence for the existence of such operators has been found, within truncations closed using (\ref{std}), in \cite{Codello:2007bd}.
\begin{figure}
\centering{}\includegraphics[scale=0.64]{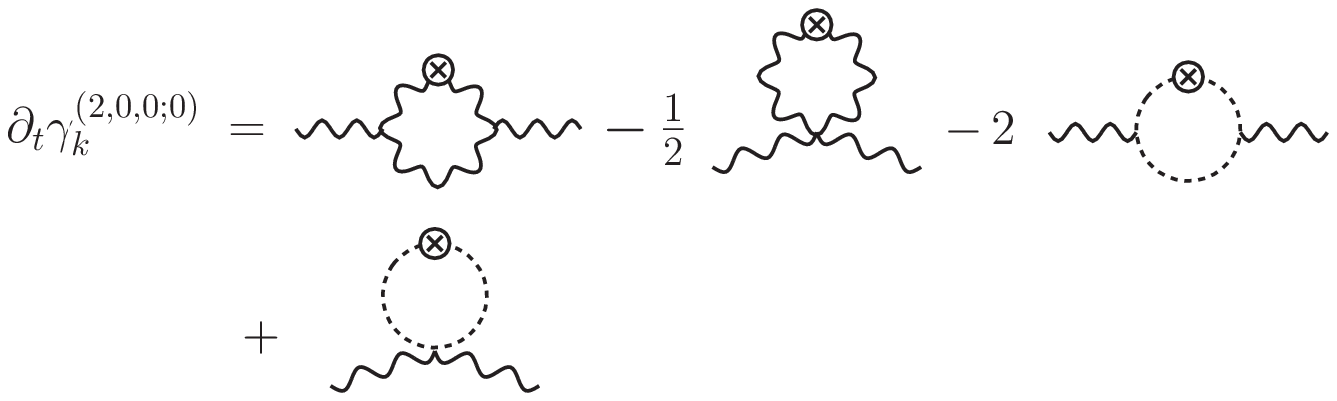}
\centering{}\includegraphics[scale=0.65]{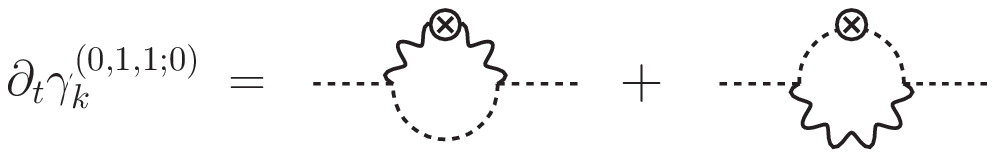}
\caption{Diagrammatic representation of the RG flow equations for the zero--field proper--vertices of the background EAA used to calculate the anomalous dimensions of the fluctuating metric and ghost fields.
Wavy lines represent the fluctuating metric, while dotted lines represent the ghosts. The cross--cap stands for a cutoff insertion.}
\end{figure}

\section{Anomalous dimensions}

The third way to close the beta functions system (\ref{BF}) is to separately calculate the anomalous dimensions
of the fluctuating metric and ghost fields that enter it. 
%
%
These can be determined as functions of $\tilde{\Lambda}_{k}$ and $\tilde{G}_{k}$ that can
successively be reinserted in the beta functions.
%
%
In doing so we make a step further in considering the flow in the enlarged theory space where the background EAA lives.
As we said, we are adopting the point of view that $\Lambda_{k}$ and $G_{k}$ are physical
couplings while $Z_{h,k}$ and $Z_{C,k}$ are not, but the influence
of these last couplings is non--trivial and it is important to account for it.

This line of reasoning has partially been implemented in \cite{Groh:2010ta,Eichhorn:2010tb} where the closure (\ref{std})
was extended by separately calculating the ghost anomalous dimension in place of setting it to zero.
The flow so obtained is similar to the standard one; in particular, it is still strongly spiraling around the non--Gaussian fixed--point.
The authors of \cite{Groh:2010ta} used a generalized heat kernel technique \cite{Benedetti:2010nr} to determine the anomalous dimension of the ghost fields;
since they use a type Ia cutoff we can compare their results directly with ours.
Another non--trivial truncation of the ghost sector has been considered in \cite{Eichhorn:2013ug}.

Our calculations of the anomalous dimensions $\eta_{h,k}$ and $\eta_{C,k}$ have been performed using the diagrammatic
techniques presented in \cite{Codello2013}, where one uses  the flow equations for the zero--field proper--vertices $\gamma^{(n,m)}_k\equiv \Gamma^{(n,m)}_k [0;0]$
of the background EAA to extract the running of the couplings. These flow equations are represented diagrammatically in Figure 1.
The full technical details of our computation will be presented elsewhere \cite{CodelloDodoricoPagani}.

Both $\eta_{h,k}$ and $\eta_{C,k}$ turn out not to depend on the cutoff operator type (i.e. on the cutoff operator used to separate fast from slow field modes) and have the following general form:
\begin{eqnarray}
&&\eta_{h,k} = \left[a_{d}(\tilde{\Lambda}_k)+c_{d}(\tilde{\Lambda}_k)\,\eta_{h,k}+e_{d}(\tilde{\Lambda}_k)\,\eta_{C,k}\right]\tilde{G}_{k}\qquad\quad\nonumber\\
&&\eta_{C,k} = \left[b_{d}(\tilde{\Lambda}_k)+d_{d}(\tilde{\Lambda}_k)\,\eta_{h,k}+f_{d}(\tilde{\Lambda}_k)\,\eta_{C,k}\right]\tilde{G}_{k}\,,\quad\label{etas}
\end{eqnarray}
where $a_{d},b_{d},c_{d},d_{d},e_{d},f_{d}$ are functions of the dimensionless cosmological constant.
Their explicit form is long and will be given in \cite{CodelloDodoricoPagani}.
Equation (\ref{etas}) constitutes a linear system for $\eta_{h,k}$ and
$\eta_{C,k}$ that can be solved to yield the anomalous dimensions
as functions solely of the physical couplings $\tilde{\Lambda}_{k}$ and $\tilde{G}_{k}$.
%
The anomalous dimensions (\ref{etas})
depend on the
(possibly scale dependent) gauge--fixing parameters that here we set to $\alpha=\beta=1$. A first study of this dependence has been made in  \cite{Eichhorn:2010tb}.
%

To solve the linear system (\ref{etas}) we rewrite it as a matrix equation:
\begin{equation}
\bar{\eta}_{k}=\left(\bar{V}+\mathbf{M}\,\bar{\eta}_{k}\right)\tilde{G}_{k}\,,
\end{equation}
where:
\begin{eqnarray}
\bar{\eta}_{k}=\left(\begin{array}{c}
\eta_{h,k}\\
\eta_{C,k}
\end{array}\right)\qquad\bar{V}=\left(\begin{array}{c}
a_{d}\\
b_{d}
\end{array}\right)\qquad\mathbf{M}=\left(\begin{array}{cc}\nonumber
c_{d} & e_{d}\\
d_{d} & f_{d}
\end{array}\right)\,.
\end{eqnarray}
The linear system is easily solved:
\begin{eqnarray}
\bar{\eta}_{k}=\tilde{G}_{k}\left(1-\tilde{G}_{k}\,\mathbf{M}\right)^{-1}\bar{V}\,;
\end{eqnarray}
or more explicitly:
\begin{widetext}
\begin{eqnarray}
\eta_{h,k}(\tilde{\Lambda}_k,\tilde{G}_{k})	&=&	\frac{a_{d} (\tilde{\Lambda}_k)[1-f_{d} (\tilde{\Lambda}_k)\tilde{G}_{k}]+b_{d}(\tilde{\Lambda}_k)e_{d}(\tilde{\Lambda}_k) \tilde{G}_{k}}{[1-c_{d}(\tilde{\Lambda}_k) \tilde{G}_{k}][1-f_{d}(\tilde{\Lambda}_k) \tilde{G}_{k}]-d_{d}(\tilde{\Lambda}_k) e_{d}(\tilde{\Lambda}_k) \tilde{G}_{k}^{2}}\tilde{G}_{k} \nonumber\\
\eta_{C,k}(\tilde{\Lambda}_k,\tilde{G}_{k})	&=&	\frac{b_{d} (\tilde{\Lambda}_k)[1-c_{d}(\tilde{\Lambda}_k) \tilde{G}_{k}]+a_{d}(\tilde{\Lambda}_k) d_{d}(\tilde{\Lambda}_k) \tilde{G}_{k}}{[1-c_{d}(\tilde{\Lambda}_k) \tilde{G}_{k}][1-f_{d}(\tilde{\Lambda}_k) \tilde{G}_{k}]-d_{d}(\tilde{\Lambda}_k) e_{d}(\tilde{\Lambda}_k) \tilde{G}_{k}^2}\tilde{G}_{k} \,. \label{etas2}
\end{eqnarray}
\end{widetext}
Inserting (\ref{etas2}) back  in the beta functions (\ref{BF}) gives the \emph{new} RG improved form of $\partial_{t}\tilde{\Lambda}_{k}$
and $\partial_{t}\tilde{G}_{k}$ that accounts for the non--trivial influence that $Z_{h,k}$ and $Z_{C,k}$ have on their flow.

Just for illustrative purposes, we report here the beta function for Newton's constant at
$\tilde{\Lambda}_{k}=0$ in $d=4$:
\begin{equation}
\partial_{t}\tilde{G}_{k}=2 \tilde{G}_{k}-\frac{\frac{11}{3 \pi }+\frac{1037}{576 \pi^2}\tilde{G}_{k}+\frac{4441}{18432 \pi ^3}\tilde{G}_{k}}{1+\frac{3 }{32 \pi }\tilde{G}_{k}+\frac{157 }{4608 \pi ^2}\tilde{G}_{k}^2}\tilde{G}_{k}^2\,,
\end{equation}
together with the anomalous dimensions is this case:
\begin{eqnarray}
\eta_{h,k}(\tilde{G}_{k})	&=&	\frac{\frac{73}{48 \pi }+\frac{2699 }{4608 \pi ^2} \tilde{G}_{k}}{1+\frac{3 }{32 \pi }\tilde{G}_{k}+\frac{157 }{4608 \pi ^2}\tilde{G}_{k}^2}  \tilde{G}_{k}\nonumber\\
\eta_{C,k}(\tilde{G}_{k})	&=&	-\frac{\frac{19 }{6 \pi }+\frac{1519 }{4608 \pi ^2}\tilde{G}_{k}}{1+\frac{3 }{32 \pi }\tilde{G}_{k}+\frac{157 }{4608 \pi ^2}\tilde{G}_{k}^2}\tilde{G}_{k}\,.
\end{eqnarray}
One can appreciate the non--perturbative nature of these results, even in this simple truncation.
\begin{figure}
\centering{}
\includegraphics[scale=0.67]{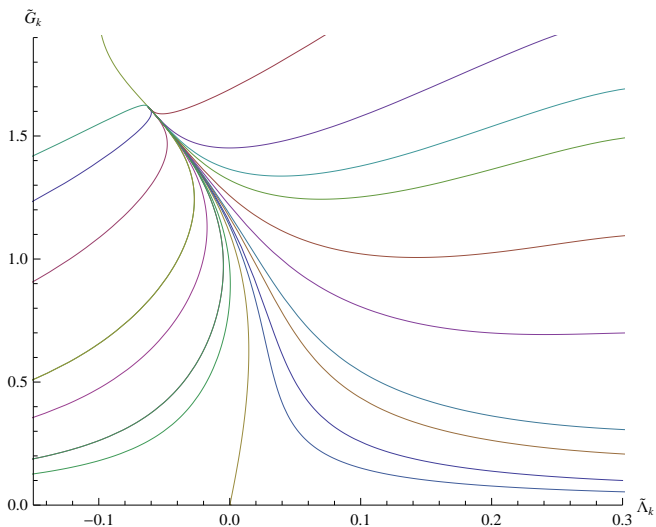}
\caption{RG flow in $d=4$ in the $(\tilde{\Lambda}_{k},\tilde{G}_{k})$
plane for the closure of the beta functions obtained by inserting back in (\ref{BF}) the independently computed anomalous dimensions $\eta_{h,k}$ and $\eta_{C,k}$.}
\end{figure}

The result of the numerical integration of these beta functions, in $d=4$, is plotted in Figure 2. 
%
Note that, despite these new beta functions differ non--trivially from the one--loop (\ref{BF1L}) and
standard RG improved (\ref{BFstd}) ones, we still find one UV attractive non--Gaussian fixed--point.
This time the critical exponents are real.
This is clearly reflected in the fact that the flow next to the 
non--Gaussian fixed--point is no more spiraling as instead was in the previous cases.
Real critical exponents are also suggested by the analysis of \cite{Benedetti:2013jk}.
%
%
The fixed--point values of the dimensionless couplings and the critical exponents are also given in Table 1.
Our fixed--point value of the dimensionless cosmological constant is negative and very small;
however, the necessary inclusion of matter contributions will in any case change its value.

%

If we insert the fixed--point values of the cosmological and Newton's
constants in (\ref{etas2}) we can determine the fixed--point values for the anomalous
dimensions $\eta_{h*}$ and $\eta_{C*}$.
The numerical values we find are reported in Table 1, together with previous estimates.
The anomalous dimension of $h_{\mu\nu}$ results positive, while the
anomalous dimension of the ghost fields is negative, as also found in \cite{Groh:2010ta,Eichhorn:2010tb}.

%


\section{Discussion}

In this letter we have shown how to account for the non--trivial influence that
the anomalous dimensions $\eta_{h,k}$ and $\eta_{C,k}$ of the fluctuating
fields have on the RG flow of the cosmological and Newton's constants.
We have derived new RG improved  beta functions for these couplings
which still exhibit  a UV attractive non--Gaussian fixed--point, but we have found real critical exponents.
These results reinforce the Asymptotic Safety scenario in quantum gravity.

%
%

The closure method proposed in this letter is general and can be applied to the beta functions
of higher derivative gravity couplings \cite{Codello:2006in},
to the beta functions obtained using the first--order formalism \cite{Tetradi}
and even to the beta functions present in non--local truncations of the gravitational EAA \cite{Satz:2010uu}.
Our method can be extended to applications of  the EAA to the renormalization of other theories with local symmetries,
as non--linear sigma models \cite{NLSM},
the theory of membranes \cite{Codello:2011yf}
or Horava--Liftshitz theories of gravity \cite{Rechenberger:2012dt}.

It is also important to understand the relation between our closure method and other 
complementary strategies proposed in the literature.
In \cite{Manrique:2009uh}
bi--metric truncations were constructed using invariants made with both $g_{\mu\nu}$ and $\bar{g}_{\mu\nu}$;
in  \cite{Pawlowski}  the problem has been studied using the Vilkovisky--DeWitt formalism;
in \cite{Christiansen:2012rx} an analysis similar to ours has been performed, where the flow of the fluctuating metric
(zero--field) two--point function in Landau gauge has been used to extract the beta functions of the cosmological and Newton's constants;
these exhibiting a non--Gaussian fixed point with real critical exponents.


\vspace*{.5cm}



\begin{thebibliography}{10}

\bibitem{Weinberg}S.~Weinberg in \emph{General Relativity, an Einstein Centenary Survey},
S.W.~Hawking and W.~Israel (Eds.) (1979) Cambridge University Press.

\bibitem{Hooft}
  G.~'t Hooft and M.~J.~G.~Veltman,
  Annales Poincare Phys.\ Theor.\ A {\bf 20} (1974) 69.

\bibitem{Reuter:1996cp}
  M.~Reuter,
  Phys.\ Rev.\ D {\bf 57} (1998) 971
  hep-th/9605030.
  
  \bibitem{Codello:2008vh}
  A.~Codello, R.~Percacci and C.~Rahmede,
  Annals Phys.\  {\bf 324} (2009) 414
  arXiv:0805.2909 [hep-th].
  
\bibitem{Codello:2007bd}
  A.~Codello, R.~Percacci and C.~Rahmede,
  Int.\ J.\ Mod.\ Phys.\ A {\bf 23} (2008) 143
  arXiv:0705.1769 [hep-th];
  P.~F.~Machado and F.~Saueressig,
  Phys.\ Rev.\ D {\bf 77} (2008) 124045
  arXiv:0712.0445 hep-th];
  K.~Falls, D.~F.~Litim, K.~Nikolakopoulos and C.~Rahmede,
  arXiv:1301.4191 [hep-th].

\bibitem{Reuter:1993kw}
  M.~Reuter and C.~Wetterich,
  Nucl.\ Phys.\ B {\bf 417} (1994) 181.
  
 \bibitem{Codello2013}
 A.~Codello, PhD. Thesis (2010), Johannes Gutenberg--Universit\"at, Mainz;
A.~Codello, arXiv:1304.2059 [hep-th].

\bibitem{Groh:2010ta}
  K.~Groh and F.~Saueressig,
  J.\ Phys.\ A {\bf 43} (2010) 365403
  arXiv:1001.5032 [hep-th].
  
\bibitem{Percacci:2005wu}
  R.~Percacci,
  Phys.\ Rev.\ D {\bf 73} (2006) 041501
  hep-th/0511177;
  A.~Codello,
  New J.\ Phys.\  {\bf 14} (2012) 015009
  arXiv:1108.1908 [gr-qc].

\bibitem{Dou:1997fg}
  D.~Dou and R.~Percacci,
  Class.\ Quant.\ Grav.\  {\bf 15} (1998) 3449
  [hep-th/9707239].
    
\bibitem{Rosten:2011mf}
  O.~J.~Rosten,
  arXiv:1106.2544 [hep-th].

\bibitem{Eichhorn:2010tb}
  A.~Eichhorn and H.~Gies,
  Phys.\ Rev.\ D {\bf 81} (2010) 104010
  arXiv:1001.5033 [hep-th].
  
\bibitem{Benedetti:2010nr}
  D.~Benedetti, K.~Groh, P.~F.~Machado and F.~Saueressig,
  JHEP {\bf 1106} (2011) 079
  arXiv:1012.3081 [hep-th].
    
\bibitem{Eichhorn:2013ug}
  A.~Eichhorn,
  arXiv:1301.0632 [hep-th].
  
  \bibitem{CodelloDodoricoPagani}
 A.~Codello, G.~D'Odorico and C.~Pagani, in preparation.
 
\bibitem{Benedetti:2013jk}
  D.~Benedetti,
  arXiv:1301.4422 [hep-th].
           

 
   \bibitem{Codello:2006in}
  A.~Codello and R.~Percacci,
  Phys.\ Rev.\ Lett.\  {\bf 97} (2006) 221301
  hep-th/0607128;
  D.~Benedetti, P.~F.~Machado and F.~Saueressig,
  Mod.\ Phys.\ Lett.\ A {\bf 24} (2009) 2233,
  arXiv:0901.2984 [hep-th];
  D.~Benedetti, P.~F.~Machado and F.~Saueressig,
  Nucl.\ Phys.\ B {\bf 824} (2010) 168
  arXiv:0902.4630 [hep-th];
  K.~Groh, S.~Rechenberger, F.~Saueressig and O.~Zanusso,
  PoS EPS {\bf -HEP2011} (2011) 124
  arXiv:1111.1743 [hep-th].

\bibitem{Tetradi}
  J.~-E.~Daum and M.~Reuter,
  Phys.\ Lett.\ B {\bf 710} (2012) 215
  arXiv:1012.4280 [hep-th];
  U.~Harst and M.~Reuter,
  JHEP {\bf 1205} (2012) 005
  arXiv:1203.2158 [hep-th];
  P.~Dona and R.~Percacci,
  Phys.\ Rev.\ D {\bf 87} (2013) 045002
  arXiv:1209.3649 [hep-th];
  J.~-E.~Daum and M.~Reuter,
 arXiv:1301.5135 [hep-th].
  
   
\bibitem{Satz:2010uu}
  A.~Satz, A.~Codello and F.~D.~Mazzitelli,
  Phys.\ Rev.\ D {\bf 82} (2010) 084011
  arXiv:1006.3808 [hep-th].
 
\bibitem{NLSM}
  A.~Codello and R.~Percacci,
  Phys.\ Lett.\ B {\bf 672} (2009) 280,
  arXiv:0810.0715 [hep-th];
  R.~Percacci and O.~Zanusso,
  Phys.\ Rev.\ D {\bf 81} (2010) 065012,
  arXiv:0910.0851 [hep-th];
  R.~Flore, A.~Wipf and O.~Zanusso,
  arXiv:1207.4499 [hep-th].
 


\bibitem{Codello:2011yf}
  A.~Codello and O.~Zanusso,
  Phys.\ Rev.\ D {\bf 83} (2011) 125021
  arXiv:1103.1089 [hep-th];
  A.~Codello, N.~Tetradis and O.~Zanusso,
  arXiv:1212.4073 [hep-th].
  
\bibitem{Rechenberger:2012dt}
  S.~Rechenberger and F.~Saueressig,
  JHEP {\bf 1303} (2013) 010,
  arXiv:1212.5114 [hep-th].

\bibitem{Manrique:2009uh}
  E.~Manrique and M.~Reuter,
  Annals Phys.\  {\bf 325} (2010) 785
  [arXiv:0907.2617 [gr-qc]].
  E.~Manrique, M.~Reuter and F.~Saueressig,
  Annals Phys.\  {\bf 326} (2011) 463
  arXiv:1006.0099 [hep-th].
  E.~Manrique, M.~Reuter and F.~Saueressig,
  Annals Phys.\  {\bf 326} (2011) 440
  arXiv:1003.5129 [hep-th].  
  
\bibitem{Pawlowski}I.~Donkin and J.~Pawlowski, arXiv:1203.4207 [hep-th].
  
\bibitem{Christiansen:2012rx}
  N.~Christiansen, D.~F.~Litim, J.~M.~Pawlowski and A.~Rodigast,
  arXiv:1209.4038 [hep-th].



\end{thebibliography}
\end{document}